\documentclass[review]{elsarticle}

\usepackage{lineno,hyperref}
\usepackage{graphicx}
\usepackage{subfig}
\usepackage{float}
\usepackage{tikz}
\usepackage{epstopdf}
\usepackage{amssymb}
\usepackage{amsmath}
\usepackage[ruled,vlined]{algorithm2e}
\usetikzlibrary{arrows.meta}
\usepackage{algpseudocode}

\usepackage{color}
\definecolor{xyhuCol}{rgb}{0.9, 0, 0.9}

\definecolor{massoudCol}{rgb}{0, 1, 0.7}

\begin{document}

\begin{frontmatter}
\title{Dual-criteria time stepping for weakly compressible smoothed particle hydrodynamics}
\author{Chi Zhang}
\ead{c.zhang@tum.de}
\author{Massoud Rezavand}
\ead{massoud.rezavand@tum.de}
\author{Xiangyu Hu \corref{mycorrespondingauthor}}
\cortext[mycorrespondingauthor]{Corresponding author. Tel.: +49 89 289 16152.}
\ead{xiangyu.hu@tum.de}
\address{Department of Mechanical Engineering, Technical University of Munich \\ 85748 Garching, Germany }

\begin{abstract}
Implementing particle-interaction configuration and time integration are performance intensive essentials of particle-based methods.
In this paper, 
a dual-criteria time-stepping method is proposed to improve the computational efficiency of the weakly-compressible smoothed particle hydrodynamic (WCSPH) method for modeling incompressible flows.
The key idea is to introduce an advection time criterion, 
which is based on fluid velocity field, 
for recreating the particle-interaction configuration.
Within this time criterion, 
several steps of pressure relaxation determined by the acoustic time criterion, based on the artificial speed of sound, can be carried out without updating the particle interaction configuration and much larger time-step sizes compared with the conventional counterpart. 
The method has shown optimized computational performance through CPU cost analysis.
Good accuracy and performance is obtained for the presented benchmarks implying  promising potential of the proposed method for incompressible flow and fluid-structure interaction simulations.
\end{abstract}

\begin{keyword}
Weakly-compressible SPH \sep Free surface flows  \sep Time integration
\end{keyword}

\end{frontmatter}
%
%
\newpage
\section{Introduction}
\label{intro}
As a fully Lagrangain particle-based method, 
smoothed particle hydrodynamics (SPH) was originally proposed by Lucy \cite{lucy1977numerical} 
and Gingold and Monaghan \cite{gingold1977smoothed} for modeling astrophysics problems.
Beside its original application \cite{springel2005cosmological,springel2010smoothed}, 
SPH has been successfully exploited in a broad variety of problems 
ranging from solid mechanics \cite{libersky1991smooth,benz1995simulations,monaghan2000sph,randles1996smoothed} 
to fluid mechanics \cite{monaghan1994simulating,colagrossi2003numerical,hu2006multi} 
and fluid-structure interactions (FSI) \cite{antoci2007numerical, marrone2011delta, han2018sph}.
Due to its Lagrangian feature, 
SPH is particularly well suited for modeling flows involving significantly varying topology 
and free surface \cite{shadloo2016smoothed,sun2017deltaplus,zhang2017generalized, ye2019smoothed, gotoh2018state}. 

In the context of modeling incompressible flows, 
the weakly-compressible SPH (WCSPH) method is widely applied 
and many studies have been devoted to portray the applicability of WCSPH 
to simulate scientific and engineering problems.
Typical examples include wave breaking or overtopping \cite{farahani2014three, shao2006simulation}, 
wave impact \cite{khayyer2008corrected, marrone2011delta}, 
violent sloshing \cite{gotoh2014enhancement,rafiee2011study}, 
as well as FSI problems \cite{antoci2007numerical}.
A well-known limitation of WCSPH is the computational efficiency, 
in particular arising when one aims at long time simulations of large scale systems 
within reasonable computing cost.
One type of techniques to alleviate this limitation 
is applying High Performance Computing (HPC) techniques 
by using Central Processing Units (CPUs),
Graphics Processing Units (GPUs) and Many Integrated Cores (MICs) systems
to accelerate the simulations 
\cite{sbalzarini2006ppm, crespo2011gpus, nishiura2015computational}.
Another type of the acceleration techniques is to modify the numerical algorithms 
to circumvent or optimize the critical computationally expensive operations 
\cite{mattson1999near, dominguez2011neighbour, winkler2018neighbour}.

SPH algorithms are based on pairwise interactions between neighboring particles
through a Gaussian-like kernel function, 
which is radial-symmetric and has compact support.
Therefore, implementing the particle-interaction configuration, i.e. 
determining the neighboring particles (by neighbor search) 
and computing corresponding kernel function values, 
is a critical aspect of the high-performance SPH solvers. 
Two different approaches, namely, cell-linked list (CLL) \cite{mattson1999near} 
and Verlet list (VL) \cite{verlet1967computer}, 
are widely used in the SPH community for implementing the particle-interaction configuration.

The CLL approach first partitions the computational domain into equisized cells,
typically with the size of the support or cut-off radius of the kernel function, 
such that a CLL is created for each cell with all particles within the cell.
Having the CLLs created, neighbor search is restricted to the nearest neighboring cells,
9 and 27 cells in two and three dimensions, respectively. 
Then, together with neighbor-searching operations, 
the CLL approach computes the kernel function values and interaction forces 
with the particles belong to these neighboring cells 
only if they are found within the cut-off radius.
Note that since the CLL approach does not store the neighboring-particle identities 
and the corresponding kernel function values, 
usually the neighbor search is carried out multiple times, 
together with the kernel function values recomputed accordingly, 
during a single time step.
For example, when the kick-drift-kick time integration algorithm is applied 
in a typical WCSPH method \cite{zhang2017weakly}, 
neighbor search and computing the kernel values 
are carried out twice for one single time-step.

The VL approach also first divides the computational domain into equisized cells 
but typically with the size larger than the cut-off radius. 
For each particle, all particles within the adjacent cells are then checked 
to create a VL containing the references to all potential neighboring particles.
In the simplest VL approach, since the cell size is the same as the cut-off radius, 
the VL contains only the real neighbors with non-vanishing kernel function values \cite{dominguez2011neighbour}.
Note that a VL may be used for multiple times without executing neighbor search 
if the particle-interaction configuration can be obtained without updating the VL.
Again, for example, in the above mentioned WCSPH method \cite{zhang2017weakly}, 
when the simplest VL approach is applied, 
a VL can be used twice for one single time step.

However, as pointed by Dominguez et al. \cite{dominguez2011neighbour}, 
the simplest VL approach does not show notable improvement on computational efficiency as the VL is only reused once.
In order to increase the reuse of VL, 
the VL approach has to use cells with considerably larger size than the cut-off radius.
Dominguez et al. \cite{dominguez2011neighbour} 
showed that with a 50\% increase of the cell size, a performance gain of 
$8\%$ is achieved when the VL is reused for $13$ times in $7$ time steps.
The reason for only incremental performance gain is that 
all the potential neighboring particles need to be checked for computing interaction force 
even only a small portion of them are real neighbors. 
This deficiency is due to the amount of potential particles scales 
to the cell size with the power of dimensions.

In this paper, we present a dual-criteria time stepping 
for the WCSPH method to improve computational efficiency 
when the simplest VL approach is employed.
The basic idea is, apart from the acoustic time criterion as in previous WCSPH methods,
to introduce an extra advection time criterion which is similar to that used in incompressible SPH 
(ISPH) method \cite{cummins1999sph}.
While the simplest VL is employed, 
such choice of time criteria leads to a dual-loop time-stepping algorithm
in which the recreation of VLs and computation of kernel function values 
are carried out in the outer advection time-step loops
and the pressure relaxation is carried out in the inner acoustic time-step loops.
Note that, in comparison to ISPH, the present method circumvents the solution of pressure Poisson equation (PPE), 
which requires global quantities for iterations 
and converged PPE solution to ensure numerical stability.
The reason for improved computational efficiency by the present method is twofold. 
First, larger time-steps compared with previous counterparts 
are permissible for the acoustic criterion. 
Second, even the VL contains only the real neighbors, 
it can be reused for many time during the acoustic time steps. 
The remainder of this paper is arranged as follows: 
Section \ref{sec:method} details the proposed time-steeping method. 
A number of validation test cases are presented and discussed 
together with computational efficiency analysis in Section \ref{sec:validation} 
and concluding remarks of the present study are summarized in Section \ref{sec:conclusion}.
%
%
\section{Method}\label{sec:method}
In the Lagrangian frame, the conservation of mass and momentum can be written as
\begin{equation} 
\begin{cases}\label{governingeq}
\frac{\text{d} \rho}{\text{d} t}  =  - \rho \nabla \cdot \mathbf{v} \\
\rho \frac{\text{d} \mathbf{v}}{\text{d} t}  =   - \nabla p +  \eta \nabla^2 \mathbf{v} + \rho \mathbf{g}
\end{cases},
\end{equation}
where $\mathbf{v}$ is the velocity, $\rho$ the density, 
$p$ the pressure, $\eta$ the dynamic viscosity, $\mathbf{g}$ the gravity  
and $\frac{\text{d}}{\text{d} t}=\frac{\partial}{\partial t} + \mathbf{v} \cdot \nabla$ stands for material derivative.
Following the weakly-compressible assumption \cite{monaghan1994simulating,morris1997modeling} 
for modeling incompressible flow, 
Eq. (\ref{governingeq}) is closed by an artificial isothermal equation of state (EoS)
\begin{equation} \label{eqeos}
p = c^2(\rho - \rho^0).
\end{equation}
Eq. \ref{eqeos} assumes that the density varies around $1 \% $ \cite{morris1997modeling} 
if an artificial sound speed of $ c = 10 U_{max}$ is employed, 
with $U_{max}$ being the maximum anticipated velocity within the flow.

\subsection{WCSPH method}
Following Refs. \cite{hu2006multi, liu2010smoothed, monaghan2012smoothed, vila1999particle},  
the WCSPH discretization of the continuity equation reads
\begin{equation}\label{eqcontinuity}
\frac{\text{d} \rho_i}{\text{d} t} = \rho_i \sum_j\frac{m_j}{\rho_j} \bold{v}_{ij}\cdot \nabla_{i} W_{ij} 
= 2\rho_i \sum_j\frac{m_j}{\rho_j}(\bold{v}_{i} - \overline{\bold{v}}_{ij}) \cdot \nabla_{i} W_{ij}.
\end{equation}
Here, $m_j$ is the particle mass, and
$ \bold{v}_{ij} = \bold{v}_i -  \bold{v}_j$ and $\overline{\bold{v}}_{ij} =(\bold{v}_i +  \bold{v}_j)/2$ 
are the relative and average velocities between particle $i$ and $j$, respectively.
$\nabla_i W_{ij}$ represents the gradient of the kernel function $W(|\bold{r}_{ij}|,h)$, 
where $\bold{r}_{ij} = \bold{r}_i - \bold{r}_j$ and $h$ is the smoothing length, 
with respect to particle $i$. 
Similarly, 
the discretization of the inviscid momentum equation gives 
\begin{equation}\label{eqmomentum}
\frac{\text{d} \bold{v}_i}{\text{d} t}  =- \sum_j m_j \bigg(\frac{p_i + p_j}{\rho_i \rho_j}  \bigg) \nabla_i W_{ij} 
= - 2\sum_j  m_j\frac{\overline{p}_{ij}}{\rho_i \rho_j}  \nabla_i W_{ij},
\end{equation}
where $\overline{p}_{ij} = ( p_i + p_j)/2$ denotes the average pressure between particle $i$ and $j$.
To stabilize the inviscid simulation, 
an artificial viscosity term \cite{monaghan2012smoothed, monaghan1983shock} can be included as
\begin{equation}\label{eqmomdissMon}
\begin{split}
\bigg( \frac{\text{d} \bold{v}_i}{\text{d} t} \bigg)^{(\nu)} & =  -\sum_j m_j \alpha \frac{h c}{\overline {\rho}} \frac{\bold{v}_{ij} \cdot \bold{r}_{ij} }{|\bold{r}_{ij}|^2}\nabla_{i} W_{ij}  
\end{split} 
\end{equation}
where $\overline{\rho} = (\rho_i + \rho_j)/2$, and $0 \leq \alpha \leq 1.0$ is a tunable parameter. 
An alternative approach to stabilize the simulation is introducing a Riemann solver 
to realize the pairwise particle interactions \cite{vila1999particle, zhang2017weakly}.
Specifically, the inter-particle averages $\overline{\bold{v}}_{ij}$ and $\overline{p}_{ij}$ 
in Eqs. (\ref{eqcontinuity}) and (\ref{eqmomentum}) are replaced by
\begin{equation} \label{eqriesolverlinear}
\overline{\bold{v}}_{ij} = U^{\ast} \bold{e}_{ij} +  ( \overline{\bold{v}}_{ij} - \overline{U}\bold{e}_{ij} ), \quad 
\overline{p}_{ij} =  P^{\ast},
\end{equation}
where $\overline{U} = \bold{v}_{ij} \cdot \bold{e}_{ij}$, 
$U^{\ast}$ and $P^{\ast}$ are the solution of inter-particle Riemann problem 
along the unit vector $\bold{e}_{ij} = -\bold{r}_{ij}/r_{ij}$ pointing form particle $i$ to particle $j$.
For viscous flows, 
Eq. (\ref{eqmomdissMon}) is replaced by the physical shear force term obtained through \cite{hu2006multi}
\begin{equation}\label{eqviscous}
\bigg( \frac{\text{d} \bold{v}_i}{\text{d} t} \bigg)^{(\nu)} 
= 2\sum_j m_j \frac{\eta}{\rho_i \rho_j} \frac{{{\mathbf{v}}_{ij}}}{{{r}_{ij}}} \frac{\partial {{W}_{ij}}}{\partial {{r}_{ij}}},
\end{equation}
where $\eta$ is the dynamic viscosity of the fluid.
The WCSPH method may suffer from particle clumping and void regions in high Reynolds number flows.
As a remedy, the transport-velocity formulation \cite{Adami2013,zhang2017generalized} with the following form
\begin{equation}\label{eqviscous}
 \frac{\text{d} \bold{\widetilde{v}}_i}{\text{d} t} 
= \frac{\text{d} \bold{v}_i}{\text{d} t} - 2\sum_j  m_j\frac{p^0}{\rho_i \rho_j}  \nabla_i W_{ij},
\end{equation}
is employable for modeling internal flows to improve the accuracy and stability.
Here, $p^0$ is the background pressure and $\bold{\widetilde{v}}$ represents the particle transport velocity, 

\subsection{Dual-criteria time stepping}
\label{sectimescheme}
The dual-criteria time-stepping employs two time-step size criteria 
characterized by the particle advection and the acoustic velocities, respectively.
The time-step size determined by the advection criterion, 
termed $\Delta t_{ad}$, 
has the following form
\begin{equation}\label{dt-advection}
\Delta t_{ad}   =  {CFL}_{ad} \min\left(\frac{h}{|\mathbf{v}|_{max}}, \frac{h^2}{\nu}\right),
\end{equation}
where $CFL_{ad} = 0.25$, 
$|\mathbf{v}|_{max}$ is the maximum particle advection velocity in the flow 
and $\nu$ the kinematic viscosity. 
Note that this criterion is the same as that used in the ISPH method \cite{hu2007incompressible}.
The time-step size according to the acoustic criterion, 
termed $\Delta t_{ac}$, 
has the form
\begin{equation}\label{dt-relax}
\Delta t_{ac}   = {CFL}_{ac} \frac{h}{c + |\mathbf{v}|_{max}},
\end{equation}
where ${CFL}_{ac} = 0.6$. 
Note that this criterion gives much larger time-step size
than that employed in conventional time integration for WCSPH simulations 
\cite{monaghan1994simulating, liu2010smoothed, monaghan2012smoothed}.

In this paper, 
while the advection criterion controls the updating frequency of the simplest VL 
and the corresponding kernel function values,
the acoustic criterion determines the frequency of the pressure relaxation process, 
i.e. the time integration of the particle density, pressure and velocity.
Accordingly, 
during one advection step, 
the pressure relaxation process is carried out approximately $k \simeq \frac{\Delta t_{ad}}{\Delta t_{ac}} $ times.
During the pressure relaxation process, 
the particle-interaction configuration is considered to be fixed in space, 
same as the ISPH method and temporally similar to the case for a Eulerian method.
As a consequence, a large $CFL_{ac} = 0.6$ number typically for a  Eulerian method, 
for example the WENO scheme \cite{jiang1996efficient}, is allowable without introducing numerical instability.

The details of the time stepping procedure are given in the following.
At the beginning of the advection step, the fluid density field of free-surface flows is reinitialized by
\begin{equation} \label{eqrhosum}
\rho_i = \max(\rho^*, \rho^0 \frac{ \sum W_{ij}}{\sum W^0_{ij}}) ,
\end{equation}
where $\rho^*$ denotes the density before re-initialization and superscript $0$ represents the initial reference value.
For flows without free surface, Eq.  \ref{eqrhosum} is merely modified as 
\begin{equation} \label{eqrhosumnosurface}
\rho_i =  \rho^0 \frac{ \sum W_{ij}}{\sum W^0_{ij}} .
\end{equation}
Similar to the previous re-initialization approaches in Refs. \cite{colagrossi2003numerical, zhang2017generalized}, 
Eq. \ref{eqrhosum} or \ref{eqrhosumnosurface} stabilizes the density 
which is updated by Eq. \ref{eqcontinuity} in the pressure relaxation process without updating the particle-interaction configuration.
Also, the viscous force is computed here and the transport-velocity formulation is applied if necessary.
Having the time-step size $\Delta t_{ad}$ calculated, 
the pressure relaxation process is repeated 
using a velocity-Verlet scheme \cite{verlet1967computer, adami2012generalized}
with the time-step size of $\Delta t_{ac}$, until the accumulated time interval is larger than $\Delta t_{ad}$. 
Here, we denote the beginning of an acoustic time step by superscript $n$, 
at the mid-point by $n + \frac{1}{2}$ and eventually at the new time-step by $n + 1$.
In the Verlet scheme, the velocity field is first updated to the mid-point value by
\begin{equation}\label{verlet-first-half}
\mathbf{v}_i^{n + \frac{1}{2}} = \mathbf{v}_i^n + \frac{1}{2}\Delta t_{ac} \big( \frac{d \mathbf{v}_i}{dt} \big)^{n}.
\end{equation}
Then particle position and density are updated to the next time step in the following form
\begin{equation}\label{verlet-first-mediate}
\begin{cases}
\mathbf{r}_i^{n + 1} = \mathbf{r}_i^{n + \frac{1}{2}} +  \Delta t_{ac} \mathbf{v}_i^{n +\frac{1}{2}} \\
\rho_i^{n + 1} = \rho_i^{n} + \frac{1}{2} \Delta t_{ac} \big( \frac{d \rho_i}{dt} \big)^{n+\frac{1}{2}}.
\end{cases}
\end{equation}
Finally, the velocity field is updated at the end of the time step as 
\begin{equation}\label{verlet-first-final}
\mathbf{v}_i^{n + 1} = \mathbf{v}_i^n +  \frac{1}{2} \Delta t_{ac} \big( \frac{d \mathbf{v}_i}{dt} \big)^{n + 1}.
\end{equation}

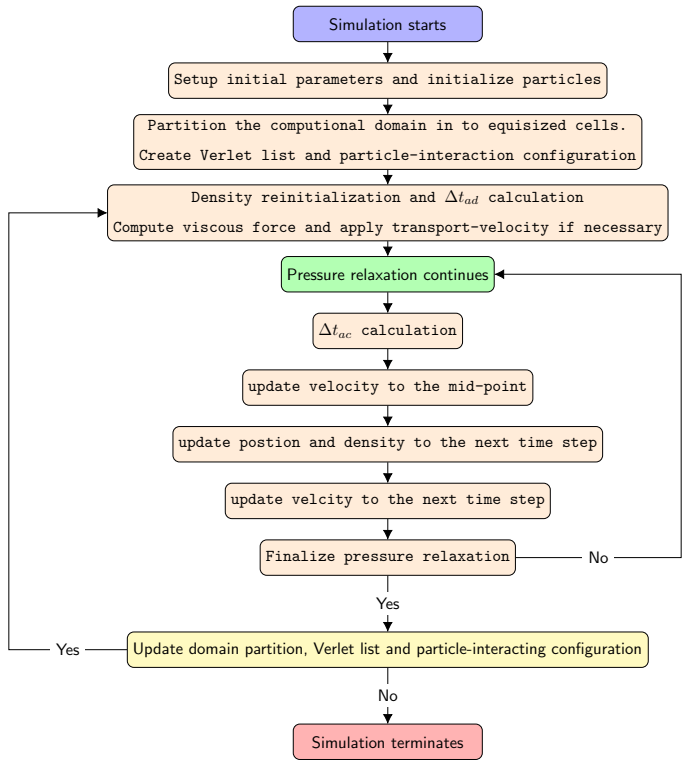
\begin{figure}[htb!]
\centering 
\tikzset{%
	>={Latex[width=2mm,length=2mm]},
	base/.style = {rectangle, rounded corners, draw=black,
		minimum width=4cm, minimum height=0.75cm,
		text centered, font=\sffamily},
	activityStarts/.style = {base, fill=blue!30},
	startstop/.style = {base, fill=red!30},
	activityRuns/.style = {base, fill=green!30},
	activityEnds/.style = {base, fill=yellow!30},
	process/.style = {base, minimum width=2.5cm, fill=orange!15,
		font=\ttfamily},
	processlarge/.style = {base, minimum width=4cm, fill=orange!15,
		font=\ttfamily},
}
\resizebox{.75\textwidth}{!}{
\begin{tikzpicture}[node distance=1.2cm, every node/.style={fill=white, font=\sffamily}, align=center]
\node (start)             [activityStarts]              {Simulation starts};
\node (onCreateBlock)     [process, below of=start]          {Setup initial parameters and initialize particles};
\node (onStartBlock)      [processlarge, below of=onCreateBlock, yshift= -0.1cm]   {Partition the computional domain in to equisized cells.\\
	 Create Verlet list and particle-interaction configuration};
\node (onResumeBlock)     [process, below of=onStartBlock, yshift= -0.3cm ]   {Density reinitialization and $\Delta t_{ad}$ calculation \\ Compute viscous force and apply transport-velocity if necessary};
\node (activityRuns)      [activityRuns, below of=onResumeBlock, yshift= -0.1cm]
{Pressure relaxation continues};
\node (onTACBlock)      [process, below of=activityRuns, yshift=0cm]
{$\Delta t_{ac}$ calculation};
\node (onPauseBlock)      [process, below of=onTACBlock, yshift=0cm]
{update velocity to the mid-point};
\node (onStopBlock)       [process, below of=onPauseBlock, yshift=0cm]
{update postion and density to the next time step};
\node (onDestroyBlock)    [process, below of=onStopBlock, yshift=0cm] 
{update velcity to the next time step};
\node (onRestartBlock)    [process, below of=onDestroyBlock, yshift=0cm]
{Finalize pressure relaxation};
\node (ActivityEnds)      [activityEnds, below of=onRestartBlock, yshift= -0.75cm]
{Update domain partition, Verlet list and particle-interacting configuration};
\node (ActivityDestroyed) [startstop, below of=ActivityEnds, yshift = -0.75cm]
{Simulation terminates};     
\draw[->]	(start) -- (onCreateBlock);
\draw[->]    (onCreateBlock) -- (onStartBlock);
\draw[->]      (onStartBlock) -- (onResumeBlock);
\draw[->]     (onResumeBlock) -- (activityRuns);
\draw[->]      (activityRuns) -- (onTACBlock);
\draw[->]      (onTACBlock) -- (onPauseBlock);
\draw[->]      (onPauseBlock) -- (onStopBlock);
\draw[->]       (onStopBlock) --  (onDestroyBlock);
\draw[->]    (onDestroyBlock) -- (onRestartBlock);
\draw[->]      (onRestartBlock) --node{Yes}(ActivityEnds);
\draw[->]     (ActivityEnds) --node{No}(ActivityDestroyed);
\draw[->] (ActivityEnds.west) --node{Yes} ++(-2.5,0) |-(onResumeBlock.west);
\draw[->] (onRestartBlock.east) --node{No} ++(3.5,0) |- (activityRuns.east);
\end{tikzpicture}
}
\caption{Flowchart of the present time stepping for the WCSPH method.}
\label{fig:flowchart}
\end{figure}

An overview of the proposed time-stepping method is shown in Fig. \ref{fig:flowchart}.
Note that the present time stepping recovers the traditional scheme 
by removing the outer loop and density reinitialization, 
applying ${CFL}_{ac} = 0.25$ and updating the VL and the kernel function values at every time step.
%
%
\section{Validation tests}\label{sec:validation}
In this section, to validate the present method,
a set of benchmark cases are simulated 
and the results are compared with those of previous experiments and simulations. 
These cases include several three-dimensional free-surface flows exhibiting violent phenomena 
such as wave impacts and breaking, 
and a two-dimensional FSI problem involving an elastic structure oscillating in flow.
The $5th$-order Wendland kernel \cite{wendland1995piecewise} 
with a smoothing length of $h = 1.3dp$, where $dp$ is the initial particle spacing, and a cut-off radius of $2.6dp$ 
is employed in all the following simulations.
As for the treatment of wall boundary, 
a one-sided Riemann solver is employed and we refer to Ref. \cite{zhang2017weakly} for more details. 

\subsection{Dam-break flow}\label{subsec31}
\begin{figure}[htb!]
	\centering
	\includegraphics[width=.85\textwidth]{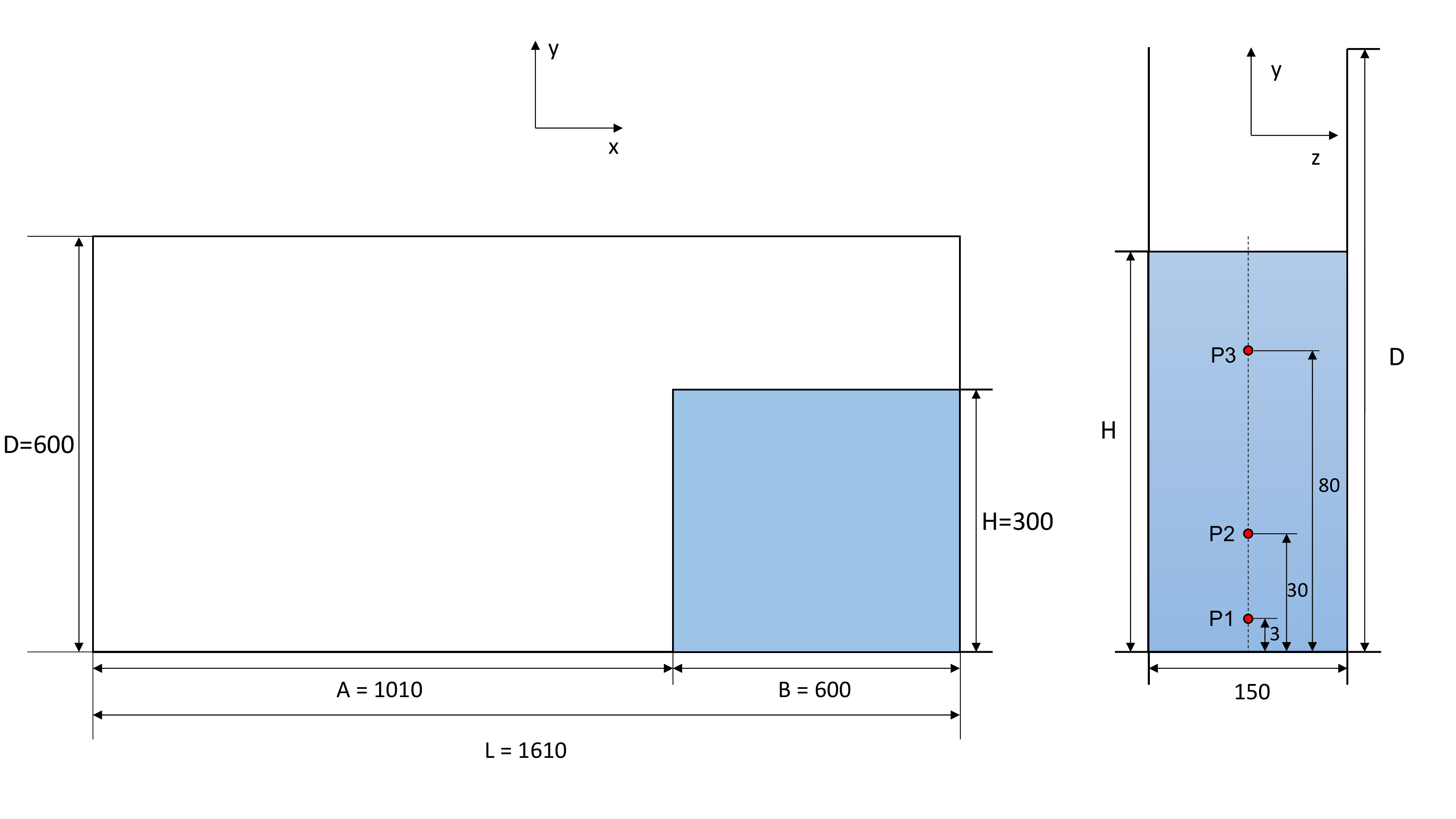}
	\caption{Schematic of the dam-break flow with the length unit of millimeter. 
		Three pressure sensors $P1$, $P2$ and $P3$ are located at the downstream wall.}
	\label{figs:damsetup}
\end{figure}

In the first case, 
we consider a dam-break flow which is widely used as a benchmark 
and has experimental data \cite{lobovsky2014experimental} available for quantitative comparison.
Following Lobovsky et al. \cite{lobovsky2014experimental}, 
the schematic and setup parameters are given in Fig. \ref{figs:damsetup}. 
We consider an inviscid flow with a density of $\rho^0 = 1000~kg / m^{3}$ 
and a gravitational constant of $g = 9.8 ~m / s^{2}$.
According to the shallow water theory, 
the maximum velocity is estimated as $v_{max} = 2\sqrt{g H}$ to set up the speed of sound, 
where $H$ is the initial water depth.
Note that, for simplicity, 
the water column starts with zero initial pressure for the computation 
instead of being released by lifting up a holding gate quickly as in the experimental setup.

\begin{figure}[htb!]
	\centering
	\includegraphics[trim = 1mm 1mm 1mm 1mm, clip,width=0.485\textwidth]{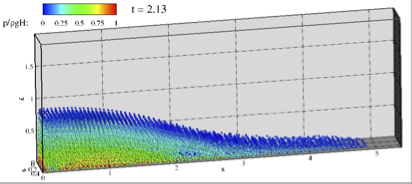}
	\includegraphics[trim = 1mm 1mm 1mm 1mm, clip,width=0.485\textwidth]{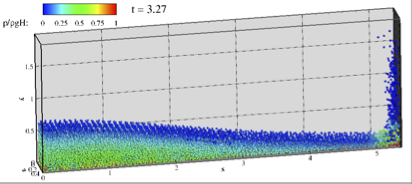}\\
	\includegraphics[trim = 1mm 1mm 1mm 1mm, clip,width=0.485\textwidth]{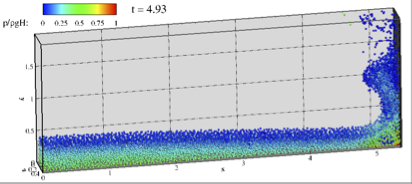}
	\includegraphics[trim = 1mm 1mm 1mm 1mm, clip,width=0.485\textwidth]{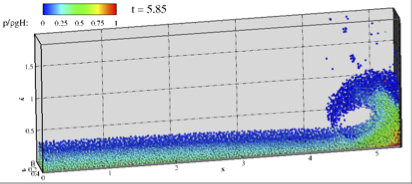}\\
	\includegraphics[trim = 1mm 1mm 1mm 1mm, clip,width=0.485\textwidth]{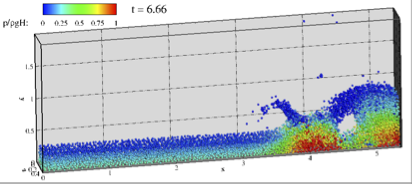}
	\includegraphics[trim = 1mm 1mm 1mm 1mm, clip,width=0.485\textwidth]{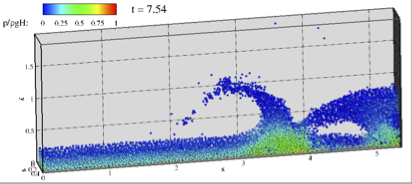}\\
	\includegraphics[trim = 2mm 1mm 1mm 1mm, clip,width=0.485\textwidth]{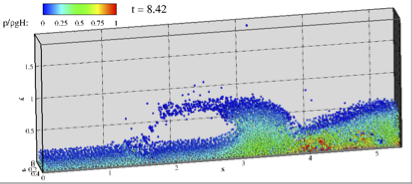}
	\includegraphics[trim = 2mm 1mm 1mm 1mm, clip,width=0.485\textwidth]{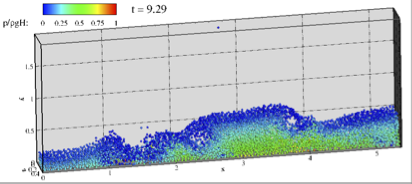}\\
	\caption{Snapshots of particle and pressure distributions during time evolution of the dam-break flow.}
	\label{figs:damfreesurface}
\end{figure}

Fig. \ref{figs:damfreesurface} illustrates several snapshots of the time evolution of free surface. 
Similar to previous numerical results \cite{ferrari2009new, zhang2017weakly}, 
the main features, viz. high roll-up along the downstream wall, 
a large reflected jet and a free surface breaking due to the re-entry of the backward wave, 
are well captured by the present method. 
\begin{figure}[htb!]
	\centering
	\includegraphics[trim = 0mm 0mm 0mm 0mm, clip,width=0.95\textwidth]{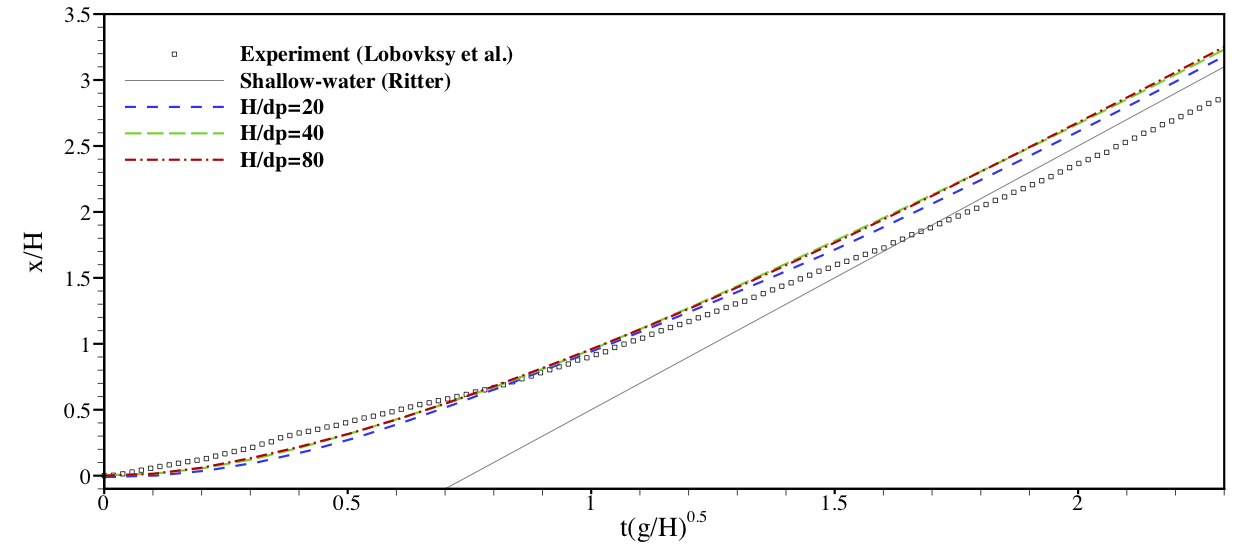}
	\caption{Time evolution of the surge-wave front in the dam-break flow.
		The experimental data are taken from Lobovsky et al. \cite{lobovsky2014experimental} 
		and the analytical solution from the shallow-water equation \cite{ritter1892fortpflanzung}.}
	\label{figs:damfront}
\end{figure}
The predicted surge-wave front with increasing spatial resolutions 
is given in Fig. \ref{figs:damfront}.
It can be observed that the surge-wave front 
converges asymptotically to the shallow-water solution at later time. 
Note that, the analytical prediction is not applicable to the initial time, 
as the assumption of shallow-water equation does not hold.
Similar to several previous simulations 
\cite{adami2012generalized, ferrari2009new, marrone2011delta, zhang2017weakly}, 
it is noted that numerically predicted surge-wave propagation is slightly faster than the experimental data, 
which is attributed to several uncertainties, 
e.g. turbulence, effect of air, wall roughness and repeatability of the experiment. 

\begin{figure}[htb!]
	\centering
	\includegraphics[trim = 0mm 0mm 0mm 0mm, clip,width=0.75\textwidth]{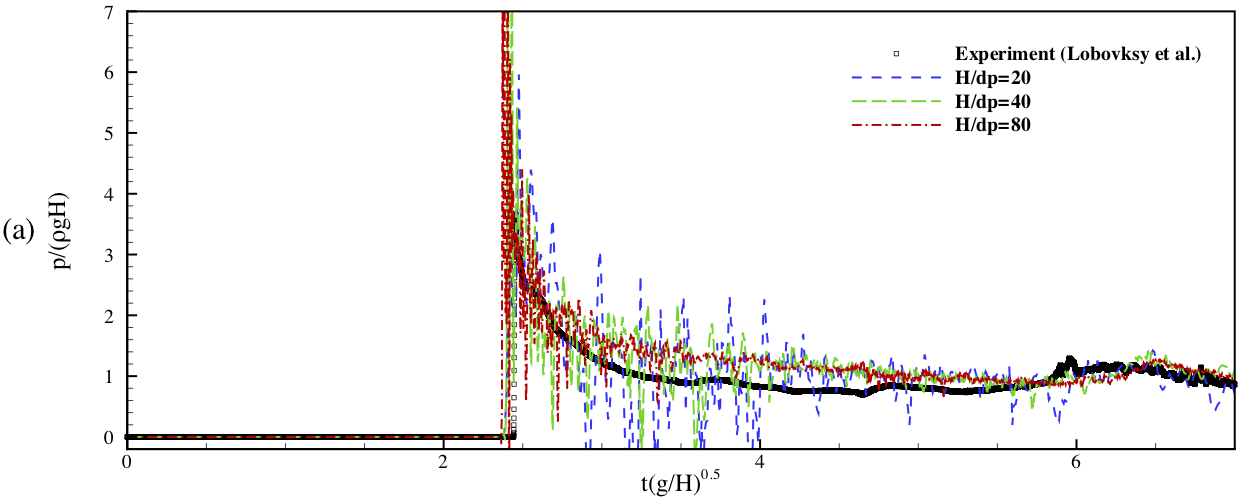}
	\includegraphics[trim = 0mm 0mm 0mm 0mm, clip,width=0.75\textwidth]{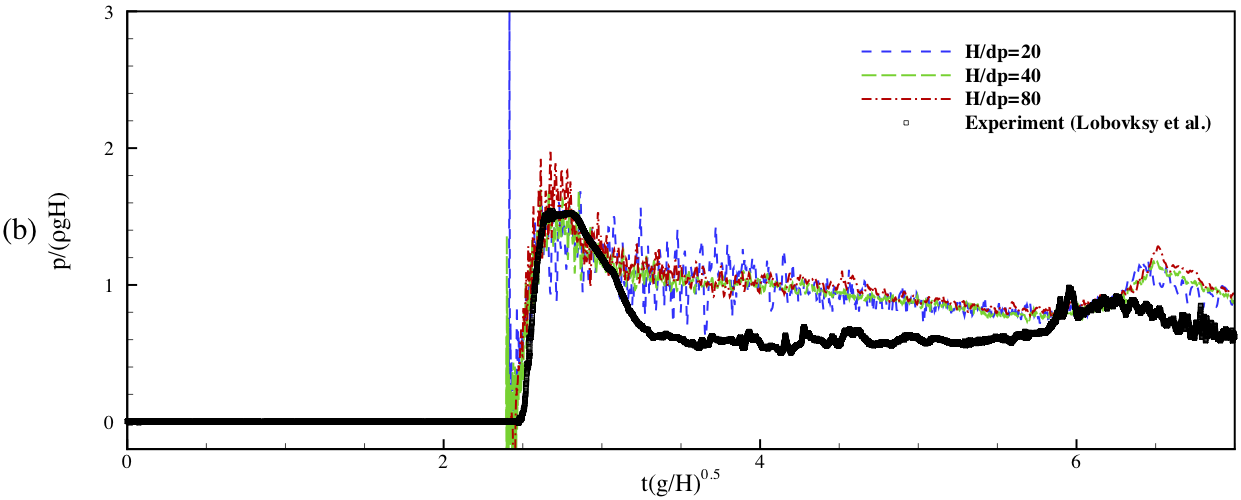}
	\includegraphics[trim = 0mm 0mm 0mm 0mm, clip,width=0.75\textwidth]{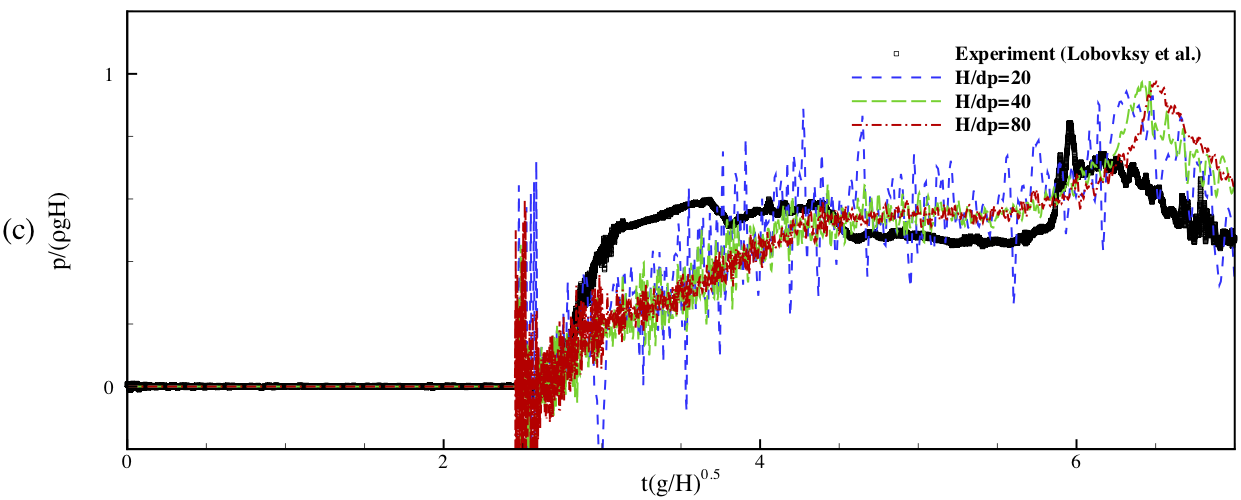}
	\caption{The time history of the pressure signals in the dam-break flow predicted at 
		$P1$ (a), $P2$ (b) and $P3$ (c) with three different spatial resolutions. 
		The experimental data is taken from Lobovsky et al. \cite{lobovsky2014experimental}.}
	\label{figs:dampressure}
\end{figure}

In the simulation, 
the measured pressure is obtained by kernel averaging from the fluid particles 
within the support radius, e.g, $2h$, of a specific sensor with
\begin{equation}\label{eqsensorpressure}
P_s =  \frac{\sum_s p_f W_{sf} }{\sum_s W_{sf} + \epsilon } ,
\end{equation}
where $\epsilon = 1.0 \times 10^{-6}$ is a small number to avoid zero denominator.
Comparison of the time history of the predicted pressure signals 
against experimental data is plotted in Fig. \ref{figs:dampressure}.
Considering that latter is post-processed by averaging the values of pressure 
on a sensor with a diameter of $d = 4.2~mm$, a good agreement is noted except for the fluctuations, 
which are decreased with increasing the spatial resolution.
Note that, the predicted pressure signal at $P3$ has lower magnitude than the experiment, 
but matches the occurrence time $t ( \sqrt{g/H} )  \sim 2.8$ quite well. 
Similar discrepancies on pressure magnitude have been observed 
in previous numerical results of Cercos \cite{cercos2015aquagpusph} 
and Zhang et al. \cite{zhang2017weakly}, simulated by different WCSPH methods. 

Here, we compare the computational efficiency of the present method 
with the traditional counterpart as interpreted in Section \ref{sectimescheme}.
The computations are carried out on an Intel(R) Xeon(R) L5520 2.27GHz Desktop 
computer with 24GiB RAM and Scientific Linux system (6.9). 
We choose the dam-break flow here because 
its has been used for performance evaluation in several previous studies \cite{winkler2018neighbour,cercos2015aquagpusph,dominguez2011neighbour}.
To analyze the performance in strong scaling behavior, 
we evaluate the CPU wall-clock time for sequential computations of 
three different spatial resolutions until the dimensionless time of $10$. 
As shown in Tab. \ref{tab:cpu-time}, 
an approximated speedup of $2.80$ is achieved by the present time-stepping method.
This result indicates that, although the VL and corresponding kernel function values 
need more memory than the conventional CLL, 
it is a good choice for large scale simulations 
when such memory requirement is not a burden.
\begin{table}[htb!]
	\caption{CPU time for simulating the dam-break flow with three spatial resolutions.}
	\label{tab:cpu-time}
	\centering
	\footnotesize{
	\begin{tabular}{|c|c|c|c|}
		\hline
		& \multicolumn{3}{c|}{Total CPU time (s)}                                                                                                                                                    \\ \hline
		\begin{tabular}[c]{@{}c@{}}Number of \\ Particles\end{tabular} & \begin{tabular}[c]{@{}c@{}}64560\\ (H/dp = 20)\end{tabular} & \begin{tabular}[c]{@{}c@{}}269472\\ (H/dp = 40)\end{tabular} & \begin{tabular}[c]{@{}c@{}}1290364\\ (H/dp = 80)\end{tabular} \\ \hline
		Traditonal                                                     & 1683.89                                                     & 29217.4                                                      & 525621                                                        \\ \hline
		Present                                                        & 583.45                                                      & 10408.2                                                      & 187314                                                        \\ \hline
		\begin{tabular}[c]{@{}c@{}}Acceleration\\ ratio\end{tabular}  & 2.88                                                        & 2.81                                                         & 2.80                                                          \\ \hline
	\end{tabular}
}
\end{table}
\subsection{Liquid sloshing}
\begin{figure}[htb!]
	\centering
	\includegraphics[width=.85\textwidth]{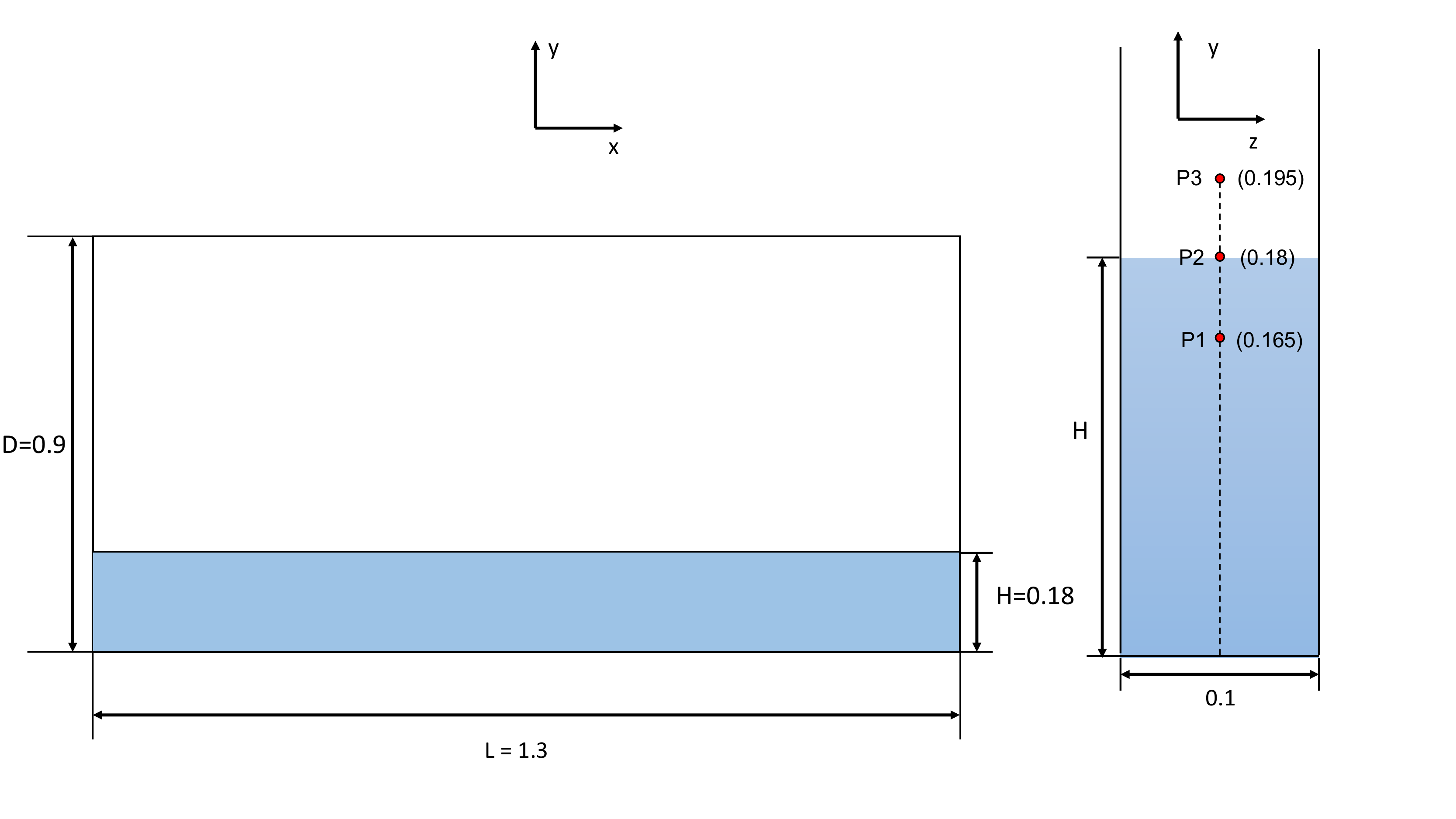}
	\caption{Schematic of the liquid-sloshing with the length unit of meter. 
	There are three pressure sensors $P1$, $P2$ and $P3$ locate at a narrow side wall as shown above. }
	\label{figs:sloshingsetup}
\end{figure}

\begin{figure}[htb!]
	\centering
	\includegraphics[trim = 0mm 1mm 1mm 0mm, clip,width=0.325\textwidth]{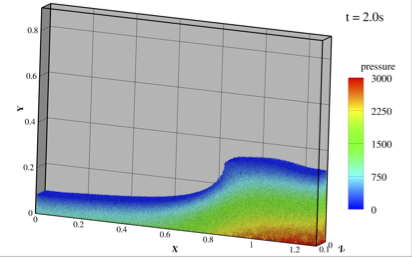}
	\includegraphics[trim = 0mm 1mm 1mm 0mm, clip,width=0.325\textwidth]{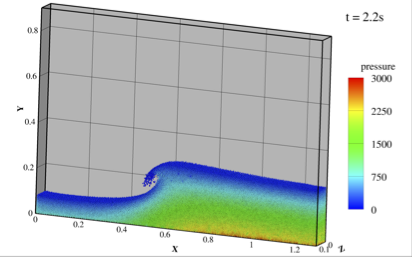}
	\includegraphics[trim = 0mm 1mm 1mm 0mm, clip,width=0.325\textwidth]{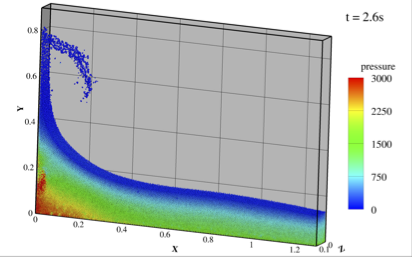}
	\caption{Snapshots of particle and pressure distributions in the simulation of liquid sloshing.}
	\label{figs:sloshingsurface}
\end{figure}

\begin{figure}[htb!]
	\centering
	\includegraphics[trim = 0mm 1mm 1mm 0mm, clip,width=0.4\textwidth]{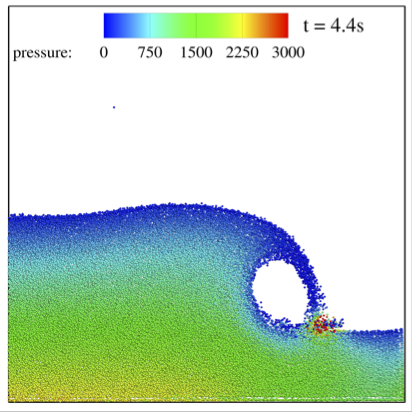}
	\includegraphics[trim = 0mm 1mm 1mm 0mm, clip,width=0.4\textwidth]{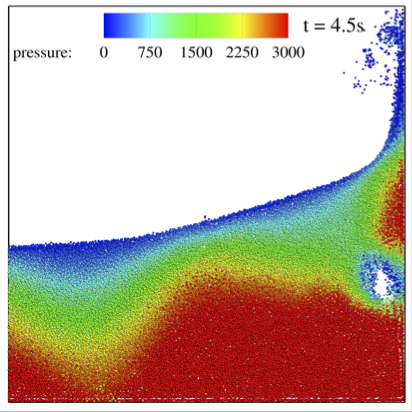}
	\caption{Zoom-in view of the free surface breaks and impacting at wall in the simulation of liquid sloshing.}
	\label{figs:sloshingsurfacefimpact}
\end{figure}

\begin{figure}[htb!]
\centering
\includegraphics[trim = 0mm 0mm 0mm 0mm, clip,width=0.9\textwidth]{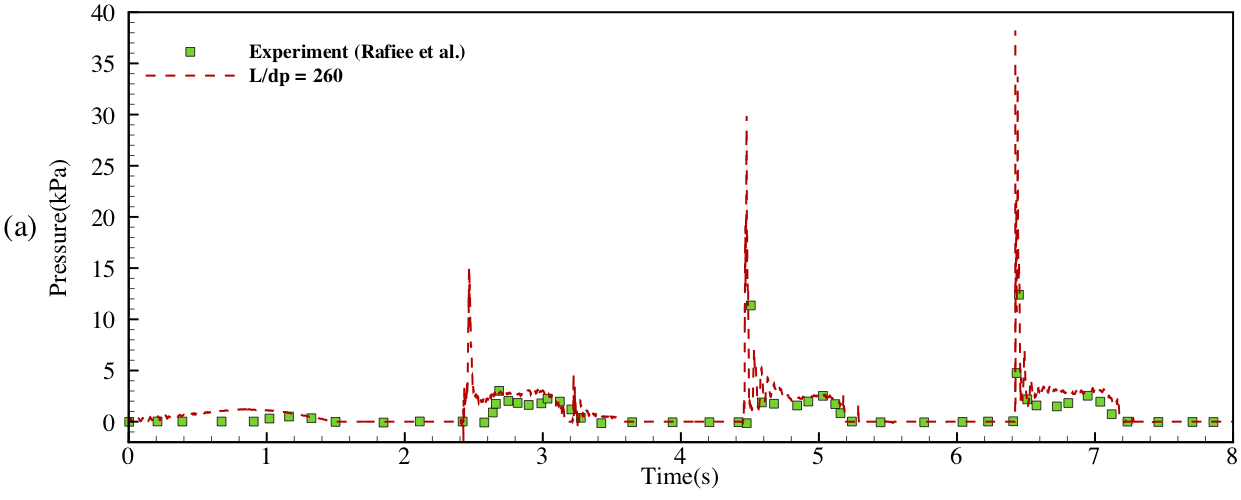}
\includegraphics[trim = 0mm 0mm 0mm 0mm, clip,width=0.9\textwidth]{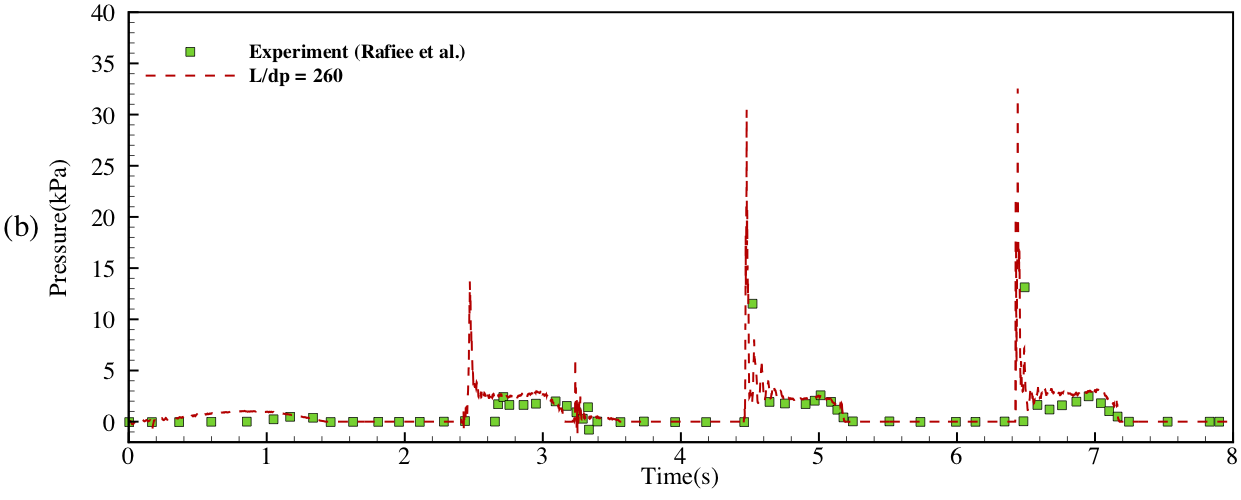}
\includegraphics[trim = 0mm 0mm 0mm 0mm, clip,width=0.9\textwidth]{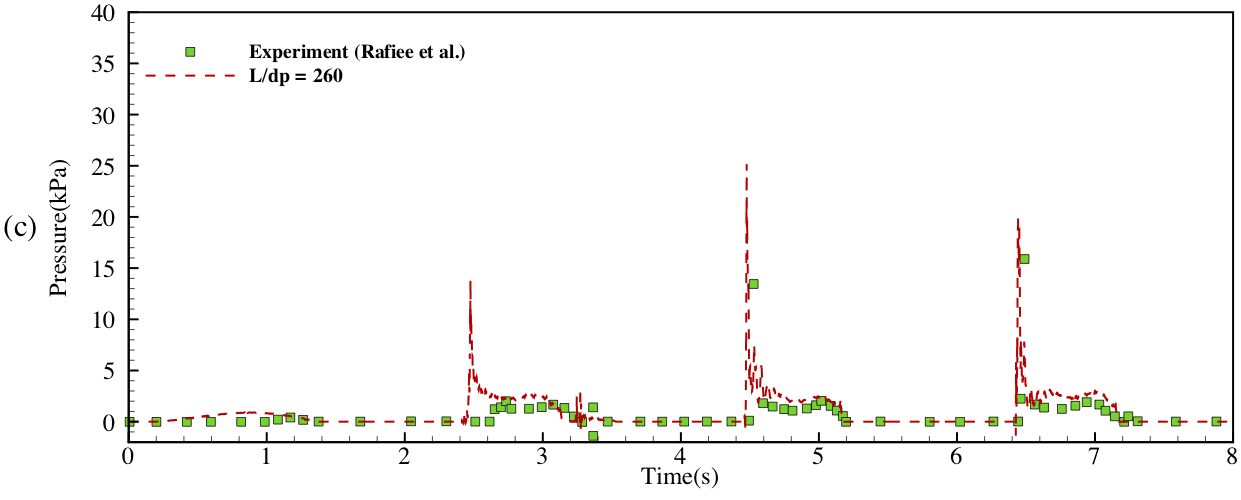}
\caption{Liquid sloshing: comparison of the time history of the impact pressure between numerical results 
	and experimental data of Rafiee et al.\cite{rafiee2011study} at pressure sensors $P1$ (a), 
	$P2$ (b) and $P3$ (c).}
\label{figs:sloshingpressure}
\end{figure}

Following Rafiee et al. \cite{rafiee2011study},
we consider an energetic violent free-surface flow, 
which involves liquid sloshing inside a rectangular tank with a low filling level. 
The geometric and setup parameters of the problem are given in Fig. \ref{figs:sloshingsetup}.
The tank motion is defined by a sinusoidal excitation in $x$-direction of 
$x = A \sin(2.0 f \pi t)$, where $A = 0.1~m$ and $f = 0.496~s^{-1}$ 
are the amplitude and frequency, respectively. 
As $f$ is close to the natural frequency of the liquid, 
thereby resulting in a highly nonlinear violent free-surface motion. 
Initially, the particles are placed on a regular lattice with a particle spacing of $dp = L/260$.
Fig. \ref{figs:sloshingsurface} shows several snapshots of particle and pressure distributions. 
The main features, e.g., a traveling wave with a crest, the resulting bore breaks 
and impacts at the wall and then forming a high run-up, are well captured by the present method.
Fig. \ref{figs:sloshingsurfacefimpact} shows the zoom-in view of the free surface breaks 
and impacting during which a void region is generated and collapsed.
This phenomenon is also observed in the experiments of Rafiee et al. \cite{rafiee2011study}, 
where the void region is filled with air.
The time history of the predicted impact pressure in comparison with experimental data 
is given in Fig. \ref{figs:sloshingpressure}.
A good agreement is noted, however, some oscillations are observed during the decay time 
and the impact pressure is overestimated.
The discrepancies may be attributed to the neglected air phase in the present simulation.
In the experimental investigations \cite{rafiee2011study}, 
there exists entrapped air pockets when the wave impacting occurs as shown in Fig. \ref{figs:sloshingsurfacefimpact}. 
The air pockets might cushion the impact and influence the peak pressure.
Note that considerably more oscillatory impact pressure is obtained 
in the results of Rafiee et al. \cite{rafiee2011study} than that predicted by the present simulation.
%
\subsection{Dam-break flow with an obstacle}
\begin{figure}[htb!]
	\centering
	\includegraphics[width=.85\textwidth]{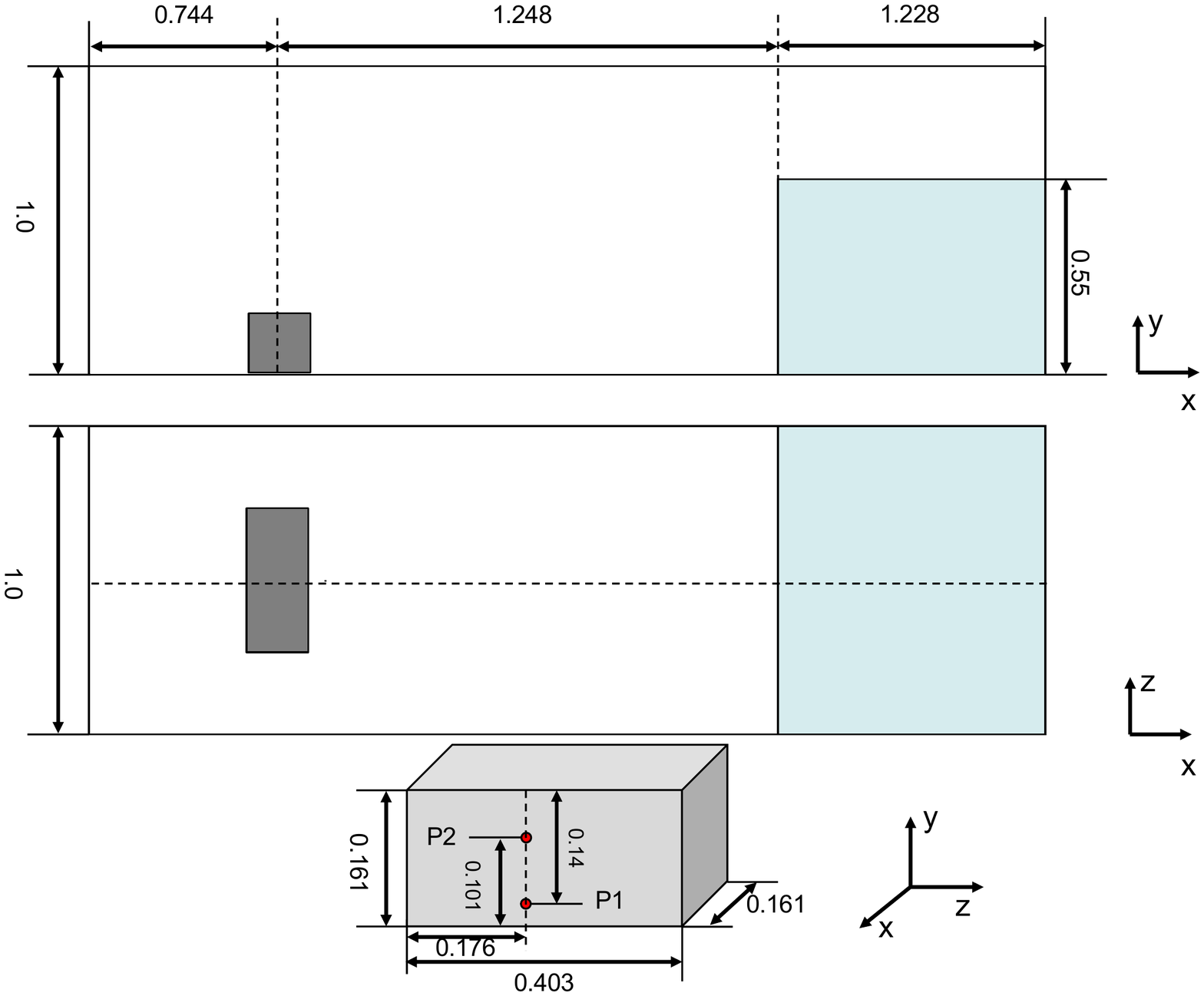}
	\caption{Schematic of the dam-break flow with an obstacle in the length unit of meter. 
	Two pressure sensors $P1$ and $P2$ are located at the front side of the obstacle.}
	\label{figs:damobsetup}
\end{figure}

We consider a dam-break flow impacting a cuboid obstacle located on the downstream horizontal bed.
This test case was first experimentally studied by Kleefsman et al. \cite{kleefsman2005volume} with 
comparison against numerical results obtained by a Eulerian volume-of-fluid (VOF) method. 
This benchmark case has been frequently used to 
validate the applicability of particle-based methods in modeling FSI problems where the structure elasticity is neglected \cite{zhang2017weakly, marrone2011delta}.
The geometric and setup parameters are briefly described in Fig. \ref{figs:damobsetup}. 
We consider an inviscid flow with a density of $\rho^0 = 1000~kg / m^{3}$ 
and the gravitational constant of $g = 9.8~m / s^{2}$. 
The particles are initially placed on a regular lattice with a particle spacing of $dp = H/55$. 

\begin{figure}[htb!]
	\centering
	\includegraphics[trim = 0mm 1mm 1mm 0mm, clip,width=0.75\textwidth]{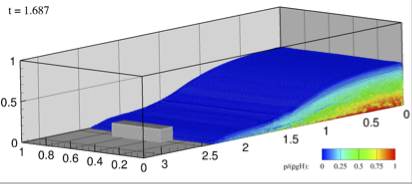}
	\includegraphics[trim = 0mm 1mm 1mm 0mm, clip,width=0.75\textwidth]{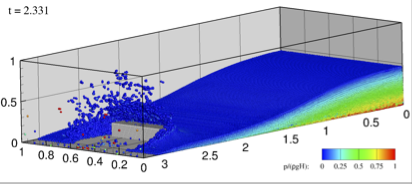}
	\caption{Snapshots of the particle and pressure distributions 
		in the simulation of the dam-break flow with an obstacle. }
	\label{figs:damobsurface}
\end{figure}

\begin{figure}[htb!]
	\centering
	\includegraphics[trim = 0mm 0mm 0mm 0mm, clip,width=0.7\textwidth]{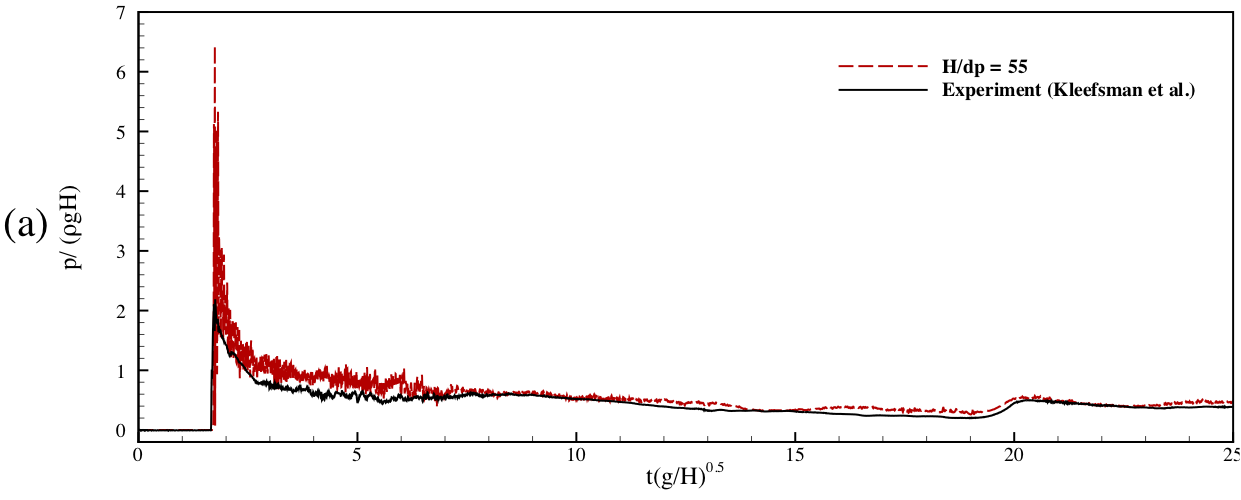}
	\includegraphics[trim = 0mm 0mm 0mm 0mm, clip,width=0.7\textwidth]{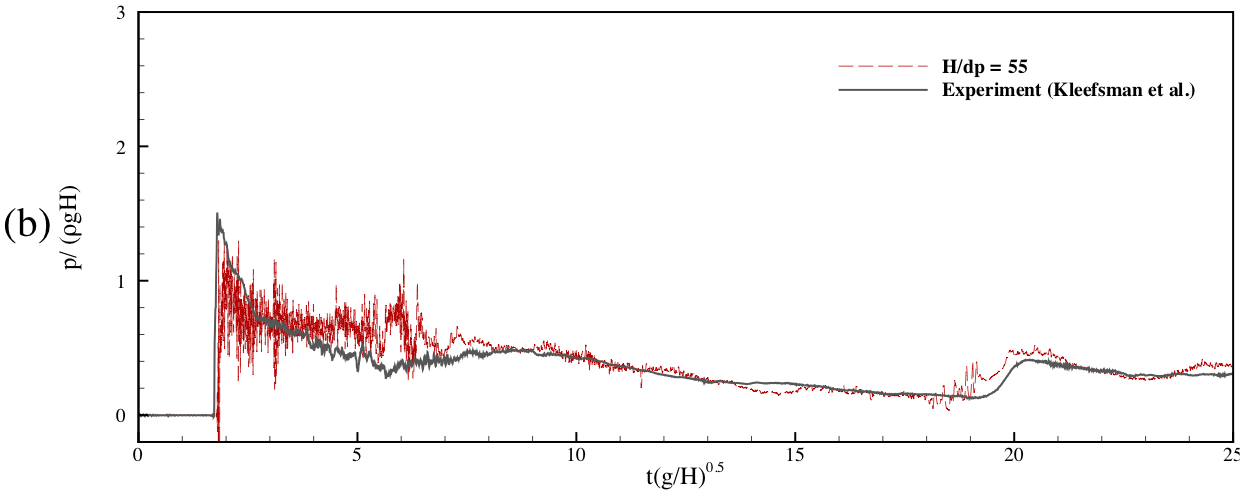}
	\caption{Dam-break flow with an obstacle: 
		time history of the pressure at $P1$ (a) and $P2$ (b) obtained from 
		the simulation and a previous experiment \cite{kleefsman2005volume}.}
	\label{figs:damobpressure}
\end{figure}

Fig. \ref{figs:damobsurface} shows several snapshots of the particle and pressure 
distributions at different time instants. 
It is observed that the free-surface profiles herein are in good agreement 
with the experimental and numerical results presented in \cite{kleefsman2005volume}, 
as well as those from previous simulations \cite{zhang2017weakly, marrone2011delta}. 
Note that, similar to the experimental observations \cite{kleefsman2005volume}, 
a splash-up is produced after the surge wave impacts at the obstacle.
Fig. \ref{figs:damobpressure} plots the time history of the pressure measured at $P1$ and $P2$, which indicate a good agreement with the experimental data except for 
the overestimated peak magnitude and the oscillations during pressure decay. 
These discrepancies are deemed to be reasonable due to the weakly-compressible formulation 
and the inviscid model employed in the present simulation.
%
\subsection{Flow-induced vibration of an elastic beam attached to a cylinder}
\begin{figure}[htb!]
	\centering
	\includegraphics[trim = 4cm 5cm 4cm 5cm, clip,width=\textwidth]{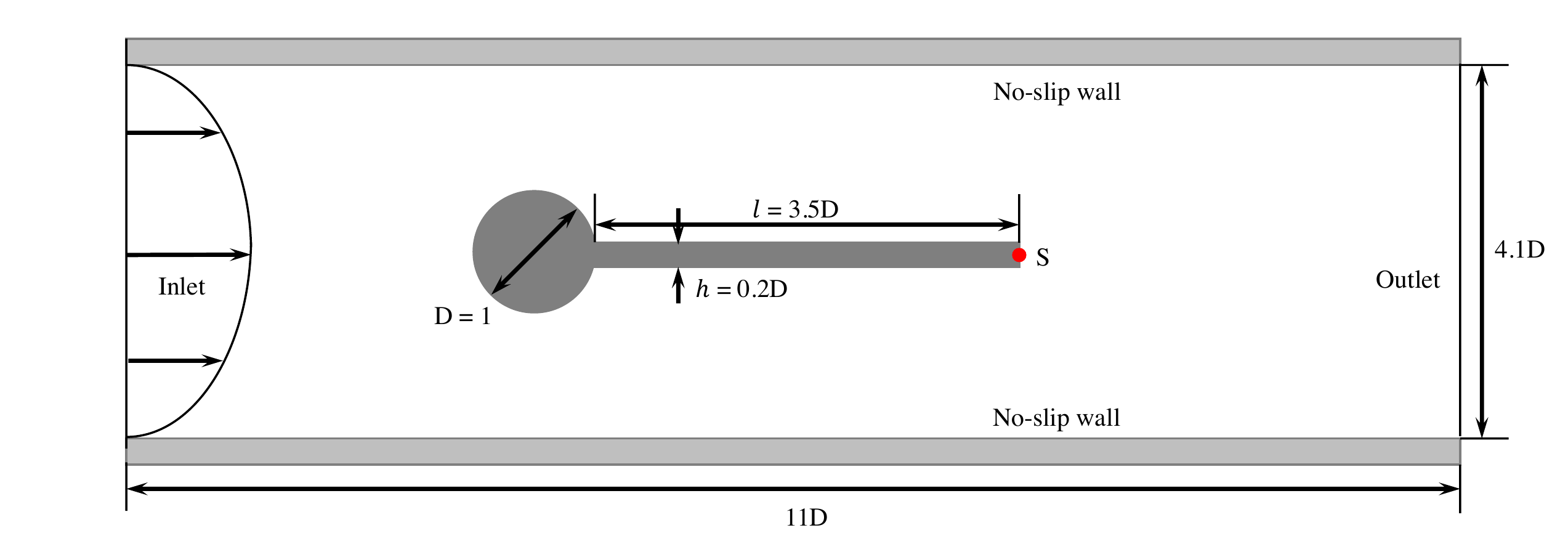}
	\caption{Schematic of the two-dimensional flow-induced vibration of a flexible beam attached to a rigid cylinder.
		The cylinder is centered at $(2D, 2D)$ measured from the left bottom corner of the computational domain.}
	\label{figs:fsisetup}
\end{figure}

Following Turek and Hron \cite{turek2006proposal} and Han and Hu \cite{han2018sph} 
a two dimensional FSI benchmark involving flow-induced vibration of 
an elastic beam attached to a rigid cylinder is considered here.
The geometric and setup parameters of the problem are shown in Fig. \ref{figs:fsisetup}. 
The density ratio of the structure to the fluid is $\rho_s/\rho_f = 10$, 
the Reynolds number $Re = \rho_{f}U_0 D/\nu = 100$, 
the dimensionless Young's modulus $E^* = E/\rho_{f}U_0^2 = 1.4 \times 10^3$ 
and the corresponding Poisson ratio $\lambda = 0.4$.
The initial particle spacing is $dp = 0.05 D$ 
and the artificial speed of sound is set to $c_0  = 20  {U}_0$.

Note that the transport-velocity formulation is applied herein with the background pressure $P_0 = \rho_0  c^2$.
No-slip boundary condition is imposed on the top and bottom walls,
while inflow and outflow conditions are applied at the left 
and right side of the domain, respectively. 
The inflow condition is given by the parabolic $x$-component velocity profile $U(t,y) = 6 {U}_0 y(H-y)/H^2$ 
with a smoothing start technique as in Refs. \cite{turek2006proposal,han2018sph}, 
where $U_0 = 1 ms^{-1}$. 
The inflow and outflow boundary conditions are implemented by 
the combination of a standard periodic boundary condition 
and a buffer inflow region with the size of 20 particle spacing by velocity relaxation.
After each acoustic time step, 
within the buffer region,  
the particle velocity is relaxed to the target value 
according to the inflow profile by
\begin{equation}\label{inflow}
\mathbf{v}_i \leftarrow \mathbf{v}_i \alpha +  \mathbf{v}(\mathbf{r}_i)(1-\alpha),
\end{equation}
where $\mathbf{v}(\mathbf{r}_i)$ is the prescribe velocity  
and $\alpha = 0.1$ is the relaxation strength chosen here.    
For the details of the SPH-discretized structure equations, 
please refer to Ref. \cite{han2018sph}. 

\begin{figure}[htb!]
	\centering
	\includegraphics[trim = 0mm 0mm 0mm 0mm, clip,width=0.8\textwidth]{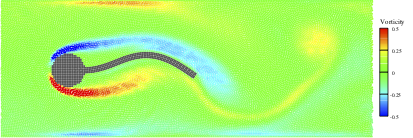}
	\includegraphics[trim = 0mm 0mm 0mm 0mm, clip,width=0.8\textwidth]{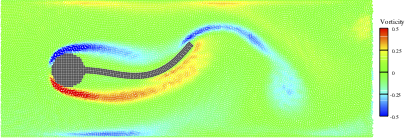}
	\includegraphics[trim = 0mm 0mm 0mm 0mm, clip,width=0.8\textwidth]{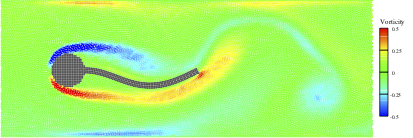}
	\includegraphics[trim = 0mm 0mm 0mm 0mm, clip,width=0.8\textwidth]{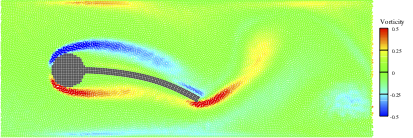}
	\caption{Flow vorticity field and beam deformation in the simulation of flow-induced vibration 
		of a flexible beam attached to a rigid cylinder. 
		From top to bottom, each panel shows the results at time instants 
		$ U_0t/D= 15.98, 16,18, 16.28$ and $16.38$, respectively.}
	\label{figs:fsi-vort}
\end{figure}

\begin{figure}[htb!]
	\centering
	\includegraphics[width=.85\textwidth]{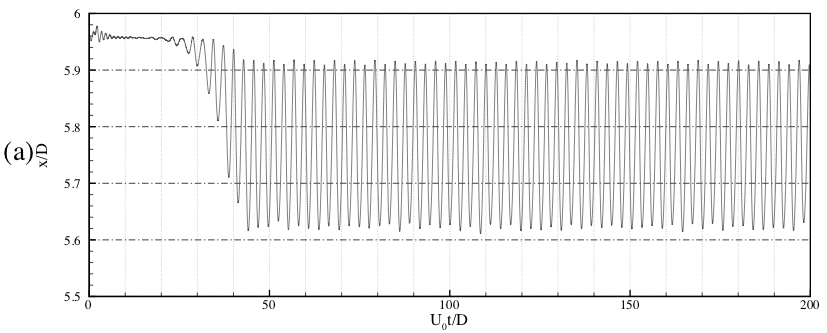}
	\includegraphics[width=.85\textwidth]{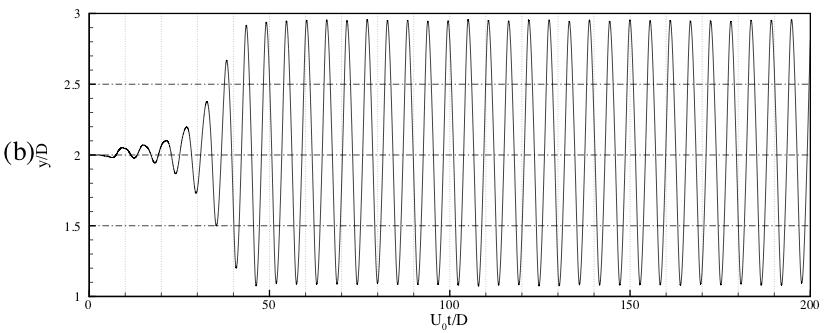}
	\includegraphics[width=.85\textwidth]{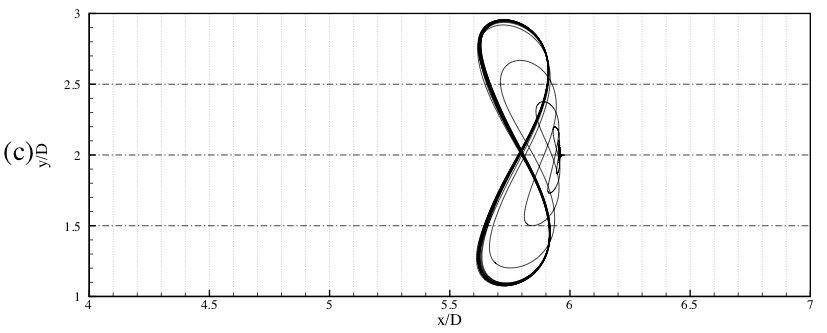}
	\caption{Flow-induced vibration of a beam attached to a cylinder: amplitude of the $x$-displacement (a), $y$-displacement (b) and the trajectory (c) of the end point $S$.}
	\label{figs:fsi-xy}
\end{figure}

Fig. \ref{figs:fsi-vort} shows the flow vorticity field and beam deformation at four different time instants 
in a typical periodic movement when self-sustained oscillation is reached.
It can be observed that these results are in quite good agreements with those from previous simulations 
\cite{turek2006proposal, han2018sph, bhardwaj2012benchmarking}. 
Fig. \ref{figs:fsi-xy} plots the temporal variation of the $x$-, $y$-displacements 
and the trajectory of the end point $S$, 
which is defined in Fig. \ref{figs:fsisetup}. 
It can be observed that the beam reaches a periodic self-sustained oscillation 
as time goes beyond a dimensionless time of 50.
A good agreement with computational results of both \cite{bhardwaj2012benchmarking} and \cite{tian2014fluid} is noted.
Note that the trajectory of point $S$ shows a typical Lissajous curve with a frequency ratio of $2:1$ 
between horizontal and vertical components \cite{bhardwaj2012benchmarking}.

Table \ref{tab:fsi-data} presents a quantitative comparison of 
the dimensionless amplitude of the oscillation in $y$-direction 
and the frequency between present computational results 
and the existing data in the literature 
\cite{turek2006proposal,bhardwaj2012benchmarking,tian2014fluid,han2018sph}. 
\begin{table}[htb!]
	\centering
	\caption{Flow-induced vibration of a beam attached to a cylinder: comparison of the present computational results against previous studies. }
	\begin{tabular}{ lcc}
		\hline
		Refs.    & Amplitude in  $y$ direction     & Frequency  \\ 
		\hline
		Turek and Hron \cite{turek2006proposal}   &$0.83$       & $0.19$      \\ 
		Bhardwaj and Mittal \cite{bhardwaj2012benchmarking}  &$0.92$       & $0.19$      \\
		Tian et al. \cite{tian2014fluid} &$0.784$ &$0.19$ \\
		Present method & $0.93$       & $0.178$     \\ 
		\hline
	\end{tabular}
	\label{tab:fsi-data}
\end{table}
The dimensionless amplitude demonstrates a good agreement, 
however, slight discrepancies are noted in the dimensionless oscillation frequency.
To evaluate the convergence of the proposed method, we performed a convergence study by decreasing the spatial resolution to $dp = 0.0125D$. 
This is found that the amplitude and the frequency, respectively, converge to $0.86$ and $0.186$, 
again, good agreement with previous results is noted.

%
%
\section{Concluding remarks}\label{sec:conclusion}
We have proposed a dual-criteria time-stepping method to optimize computational efficiency of the WCSPH method.
Apart from the acoustic time criterion which controls the pressure relaxation process,
an extra advection time criterion is introduced to control the recreation of particle-interaction configuration.
During the advection time criterion, 
several steps of pressure relaxation are processed whereas the particle-interaction configuration is frozen.
As consequence, 
significant amount of computational efforts are saved due to the fact that particle-interaction configuration can be reused for several time steps.
Another noteworthy feature of the present method is that 
a large CFL number is permissible for pressure-relaxation time-steps within the WCSPH context implying optimized computational performance.
Through computational performance analysis, 
we have demonstrated that an speedup of $2.80$ is achieved by the present method, 
in comparison to the traditional counterpart.
With preliminary examples, 
we have shown that the present method is capable of modeling free-surface flows involving violent impact and breaking events, 
as well as fluid- rigid/elastic structure interaction problems, 
with good robustness and accuracy.
While the present simulations are focusing on benchmark tests, 
the present method is expected to be employed in more complex scientific and engineering problems.
%
%
\section{Acknowledgement}
The authors gratefully acknowledge the financial support by German Research Fundation (Deutsche Forschungsgemeinschaft) DFG HU1527/10-1 and HU1527/12-1  for the present work.
%
%
\newpage
\bibliography{mybibfile}

\end{document}